\documentclass[aps,twocolumn,amsmath,amsmath,amssymb]{revtex4-2}

\usepackage{graphicx}
\usepackage{dcolumn}
\usepackage{bm}
\usepackage{float}
\usepackage[dvipsname]{xcolor}
\renewcommand{\vec}[1]{\mathbf{#1}}
\newcommand{\gvec}[1]{\boldsymbol{#1}}

\usepackage[utf8]{inputenc} 
\usepackage[T1]{fontenc}

\begin{document}

\title{Revealing Hidden Orbital Pseudospin Texture with Time-Reversal Dichroism in Photoelectron Angular Distributions}

\author{S. Beaulieu$^{1}$}
\email{beaulieu@fhi-berlin.mpg.de}
\author{J. Schusser$^{2,3}$}
\author{S. Dong$^{1}$}
\author{M. Schüler$^{4}$}
\author{T. Pincelli$^{1}$}
\author{M. Dendzik$^{1,5}$} 
\author{J. Maklar$^{1}$}
\author{A. Neef$^{1}$}
\author{H. Ebert$^{6}$}
\author{K. Hricovini$^{2,7}$}
\author{M. Wolf$^{1}$} 
\author{J. Braun$^{6}$}
\author{L. Rettig$^{1}$}
\author{J. Minár$^{3}$}
\email{jminar@ntc.zcu.cz}
\author{R. Ernstorfer$^{1}$} 
\email{ernstorfer@fhi-berlin.mpg.de}

\affiliation{
$^1$Fritz-Haber-Institut der Max-Planck-Gesellschaft, Faradayweg 4-6, 14195 Berlin, Germany \\
$^2$Laboratoire de Physique des Matériaux et Surfaces, CY Cergy Paris Université, 95031 Cergy-Pontoise, France \\
$^3$New Technologies-Research Center, University of West Bohemia, 30614 Pilsen, Czech Republic \\
$^4$Stanford Institute for Materials and Energy Sciences (SIMES), SLAC National Accelerator Laboratory, Menlo Park, CA 94025, USA \\
$^5$Department of Applied Physics, KTH Royal Institute of Technology, Hannes Alfvéns väg 12, 114 19 Stockholm, Sweden \\
$^6$Department Chemie, Ludwig-Maximilians-Universität München, Butenandtstrasse 11, 81377 München, Germany\\
$^7$LIDYL, CEA, CNRS (UMR 9222), Université Paris-Saclay, CEA Saclay, F-91191, Gif-sur-Yvette Cedex, France}

\begin{abstract}
We performed angle-resolved photoemission spectroscopy (ARPES) of bulk 2H-WSe$_2$ for different crystal orientations linked to each other by time-reversal symmetry. We introduce a new observable called time-reversal dichroism in photoelectron angular distributions (TRDAD), which quantifies the modulation of the photoemission intensity upon effective time-reversal operation. We demonstrate that the hidden orbital pseudospin texture leaves its imprint onto TRDAD, due to multiple orbitals interference effects in photoemission. Our experimental results are in quantitative agreement with both tight-binding model and state-of-the-art fully relativistic calculations performed using the one-step model of photoemission. While spin-resolved ARPES probes the spin component of entangled spin-orbital texture in multiorbital systems, we unambiguously demonstrate that TRDAD reveals its orbital pseudospin texture counterpart.   
\end{abstract}

\date{\today}
\maketitle

Locking between spin and valley degrees of freedom emerges in solids possessing a combined broken inversion symmetry and strong spin-orbit coupling, leading to peculiar valley dependent spin texture in momentum-space. This spin-valley locking leads to optical selection rules allowing for the generation of spin- and valley-polarized excited carriers \cite{Mak12,Zeng12,Bertoni16}, which can be used for all-optical selective spin injection \cite{Gmitra15,Avsar17}. In multiorbital systems, additional locking between crystal momentum and orbital degree of freedom emerges as a consequence of band hybridization, leading to complex entangled spin-orbital textures, as predicted in some topological insulators (TIs) \cite{Zhang13,Zhu13}, two-dimensional electron gases (2DEGs) \cite{King14} and transition metal dichalcogenides (TMDCs). The resulting momentum-space orbital texture can lead to orbital Hall effect (OHE) \cite{Go18}, orbital Rashba effect \cite{Park12}, and the emergence of orbital Hall insulating phases \cite{Canonico20}. Orbitronics \cite{Bernevig05}, \textit{i.e.}~encoding (quantum) information in the orbital degree of freedom, can be seen as a newly emerging field, in analogy to spin- and valleytronics \cite{Phong19,Bhowal20}. 

TMDC monolayers are emblematic materials with entangled spin, orbital and valley degrees of freedom. In a minimal electronic structure model of TMDC monolayers \cite{Liu13}, the valence band at the Brillouin zone boundary (K/K' points) can be described by $| \psi_{\textbf{k}}^{K/K'} \rangle \approx \left [ (C_0(\textbf{k})| d_{z^2}\rangle + C_{\pm 2}(\textbf{k})| d_{\pm2}\rangle) \otimes|\!\uparrow\!/\!\downarrow \rangle \right ]_{VB1} + \left [ (C_0(\textbf{k})|d_{z^2}\rangle + C_{\pm 2}(\textbf{k})| d_{\pm2}\rangle) \otimes|\!\downarrow\!/\!\uparrow \rangle \right ]_{VB2} $, where the label VB1 and VB2 represented the two first spin-orbit split valence bands and where $| d_{\pm2} \rangle = [| d_{x^2-y^2} \rangle\pm i| d_{xy} \rangle]/ \sqrt{2}$. While the spin texture is determined by the momentum-dependent spin state, the orbital texture is set by the momentum-dependent orbital pseudospin, defined as $\sigma_{i}^{K/K'}(\textbf{k}) = \langle \psi_{\textbf{k}}^{K/K'} |\hat{\sigma}_{i}| \psi_{\textbf{k}}^{K/K'} \rangle$, where $\hat{\sigma}_i$ ($i= x, y, z$) is the Pauli operator~\footnote{The Pauli matrices act on the orbital space of $\{ |d_0\rangle, |d_{\pm 2}\rangle \}$ at K/K'.}, and where $x,y$ are in-plane and $z$ out-of-plane coordinates (normal to the surface).

K and K' valleys are related to each other via the time-reversal operator, \textit{i.e}  $\hat{T}| \psi_{\textbf{k}}^{K} \rangle=| \psi_{\textbf{k}}^{K'} \rangle$. In TMDC monolayers, swapping valley indexes (time-reversal) thus reverses both the spin and orbital textures. In bulk-TMDC of 2H polytype, the adjacent layers are rotated by 180$^{\circ}$ with respect to each other, leading to opposite and alternating local spin polarization and orbital texture between neighboring layers. This peculiar layered structure naturally introduces the concept of "hidden" spin and orbital texture, which exists within each layer, but vanishes in the bulk, \textit{i.e.}~when the inversion-symmetry of the crystal is restored \cite{Zhang14}. Probing such "hidden" physical properties is experimentally challenging. While spin- and angle-resolved photoemission spectroscopy is now a well established surface sensitive technique to investigate hidden spin texture \cite{Riley14,Razzoli17,Tu20}, an experimental technique allowing to selectively and directly probe hidden orbital texture has not been established. 

In this Letter, we perform extreme ultraviolet (XUV) angle-resolved photoemission spectroscopy (ARPES) of bulk 2H-WSe$_2$ for crystal orientations rotated by $\alpha$ = 60$^{\circ}$ with respect to each others ($\mathcal{R}_{60^{\circ}}$) effectively acting as the time-reversal operator ($\hat{T}$): $\mathcal{R}_{60^{\circ}}|\psi_{\textbf{k}}^{K}\rangle \equiv \hat{T}| \psi_{\textbf{k}}^{K}\rangle=| \psi_{\textbf{k}}^{K'}\rangle$. We introduce a novel observable, \textit{time-reversal dichroism in photoelectron angular distributions} (TRDAD), which probes the modulation of the photoemission intensity upon time-reversal, and which is shown to be sensitive to the orbital pseudospin texture, due to multiple orbitals interference effects in photoemission. We show that TRDAD is free of any spurious contribution from experimental geometry, which typically complicates the interpretation of the (linear or circular) dichroism in ARPES. Our experimental results are in quantitative agreement with state-of-the-art fully relativistic Korringa-Kohn-Rostoker (KKR) \textit{ab initio} calculations performed using the one-step model of photoemission \cite{Ebert11,Braun18} and with a tight-binding model, which allows us to investigate the microscopic origins of TRDAD. While we introduce this novel observable using the emblematic bulk 2H-WSe$_2$ crystal, our conclusions are fully general and TRDAD can be used to probe (hidden) orbital texture in any multiorbital systems. 

\begin{figure}
\begin{center}
\includegraphics[width=8.5 cm,keepaspectratio=true]{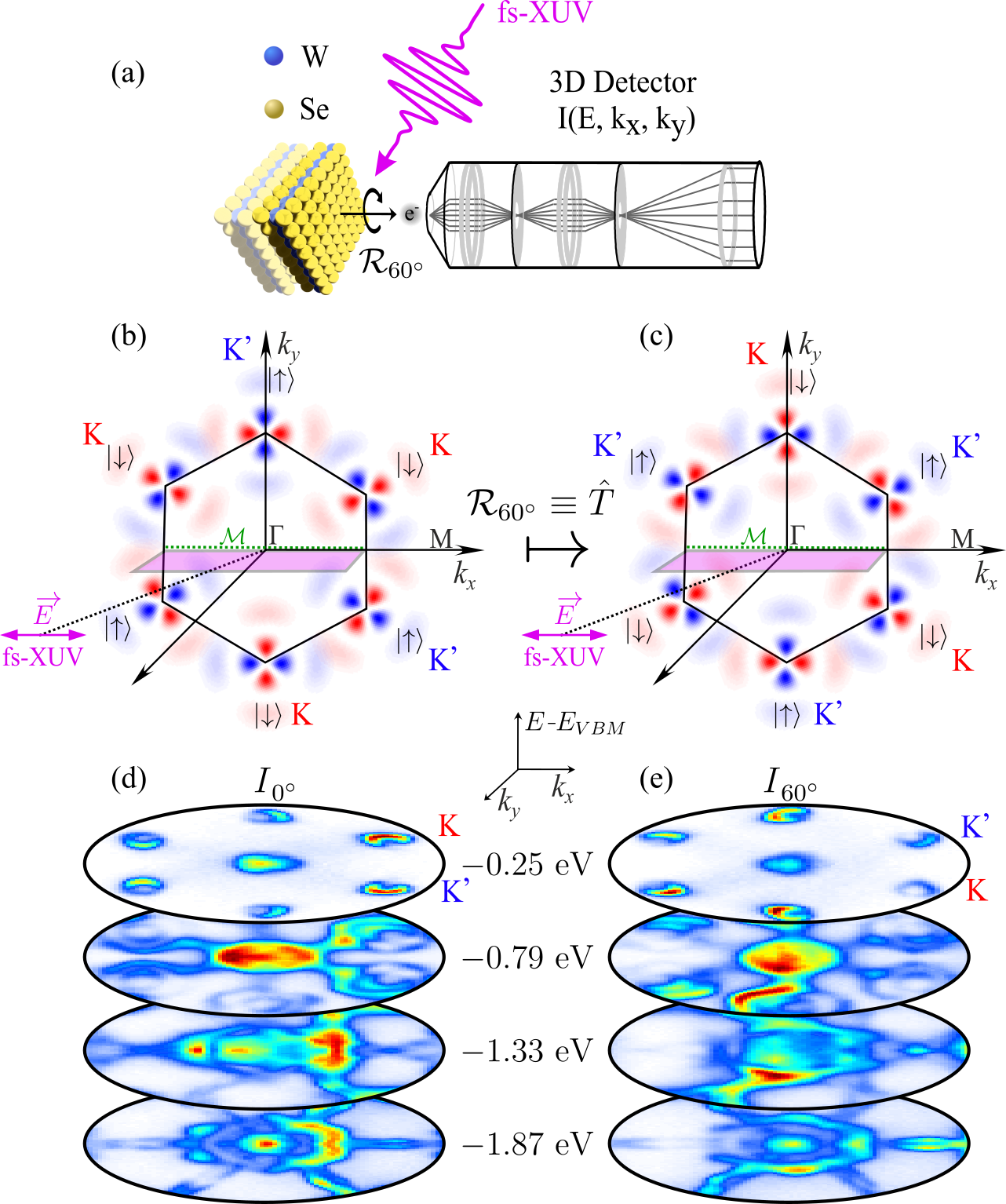}
\caption{\textbf{Modulation of the photoemission intensity upon time-reversal ($\hat{T}$)}: \textbf{(a)} Scheme of the experimental setup: a $p$-polarized fs-XUV (21.7~eV) pulse is focused onto a bulk 2H-WSe$_2$ crystal at an angle of incidence of 65$^{\circ}$ with respect to the surface normal, ejecting photoelectrons which are detected by a time-of-flight momentum microscope. \textbf{(b)-(c)} Scheme of the experimental geometry and spin-orbital texture: the scattering plane (light purple plane) coincides with the crystal mirror plane ($\mathcal{M}$, green dashed line), which is along the $\mathrm{\Gamma}$-M high symmetry direction. The spin-orbital texture is schematically represented by the orbitals and the up/down spin state at each K/K' valleys. A 60$^{\circ}$ azimuthal rotation of the crystal ($\mathcal{R}_{60^{\circ}}$) yields the transformation of K to K' valley (and vice-versa), and is analogue to the action of the time-reversal operator; \textit{i.e.} $\mathcal{R}_{60^{\circ}}|\psi_{\textbf{k}}^{K} \rangle \equiv \hat{T}|\psi_{\textbf{k}}^{K} \rangle=|\psi_{\textbf{k}}^{K'}\rangle$. \textbf{(d)-(e)} Constant energy contours for different energies, $\mathrm{E-E_{VBM}}$, measured for the two different crystal orientations described above ($I_{0^{\circ}}$ and $I_{60^{\circ}}$). The radius of each constant energy contour corresponds to 1.6~{\AA}$^{-1}$.}
\label{fig1}
\end{center}
\end{figure}

The experimental apparatus features a table-top femtosecond (fs) XUV (21.7~eV, 110~meV FWHM bandwidth) beamline \cite{Puppin19} coupled to a time-of-flight momentum microscope spectrometer (METIS~1000, SPECS GmbH), see Fig.~\ref{fig1}(a). This detector allows for simultaneous detection of the full first Brillouin zone, over an extended binding energy range, without the need to rearrange the sample geometry \cite{Medjanik17}. More details about the experimental setup can be found elsewhere \cite{Puppin19,Beaulieu20,Maklar20} and in the SM. As shown in Fig.~\ref{fig1} (b)-(e), we recorded the 3D photoemission intensity for two different crystal orientations, rotated by 60$\mathrm{^{\circ}}$ with respect to each other. Looking at the experimentally measured constant energy contours (CECs) for energy $\mathrm{E-E_{VBM}}$ = -0.25~eV, one can see that the photoemission intensity is strongly anisotropic around each K/K' valley, describing "croissant"-shaped patterns. This has been recently explained as originating from interference between photoelectrons emitted from the transition metal $d$-type orbitals \cite{Rostami19}. The azimuthal variation of the photoemission intensity around K/K' points, \textit{i.e.}~the orientation of the "croissant", changes upon rotation of the crystal by 60$\mathrm{^{\circ}}$. A modification of the momentum-resolved photoemission intensity upon time-reversal can also be seen for larger binding energies in Fig.~\ref{fig1} (d)-(e).
 
\begin{figure}
\begin{center}
\includegraphics[width=8.5 cm,keepaspectratio=true]{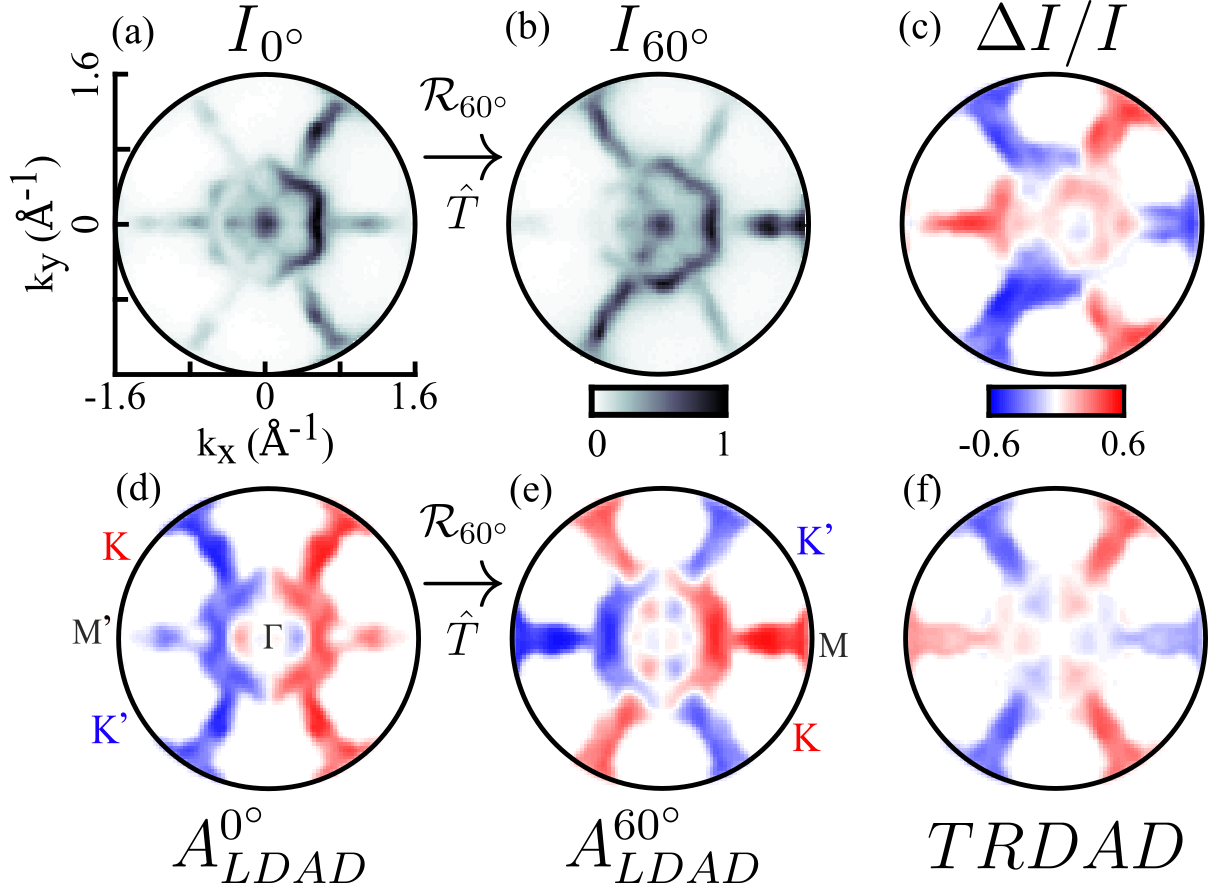}
\caption{\textbf{Extraction of the time-reversal dichroism in photoelectron angular distributions (TRDAD)}: \textbf{(a)-(b)} $I_{0^{\circ}}$ and $I_{60^{\circ}}$, the constant energy concours (CECs) for $\mathrm{E-E_{VBM}}$ = -1.60~eV, measured for two crystal orientations rotated by 60$^{\circ}$ with respect to each others. \textbf{(c)} $\Delta I/I$, the raw normalized difference, \textit{i.e.} ($I_{0^{\circ}}$ - $I_{60^{\circ}}$)/($I_{0^{\circ}}$ + $I_{60^{\circ}}$) between CECs shown in (a) and (b). \textbf{(d)-(e)} $A_{LDAD}^{0^{\circ}}$ and $A_{LDAD}^{60^{\circ}}$, the “left-right asymmetries”, reflect the photoemission intensity asymmetry between the $k_x\!<$0 and $k_x\!>$0 hemispheres, for two crystal orientations, respectively, and are calculated using Eq.~\ref{Eq:A_LDAD}. \textbf{(f)} TRDAD (calculated using Eq.~\ref{Eq:A_LDAD_T}) represents the component of $A_{LDAD}^{0^{\circ}/60^{\circ}}$ which is antisymmetric upon time-reversal (\textit{i.e.} upon 60$^{\circ}$ azimuthal rotation of the crystal).}
\label{fig2}
\end{center}
\end{figure}

Dichroism in the angular distribution (both linear, LDAD, and circular, CDAD) are powerful quantities relying on the modulation of the photoemission transition dipole matrix element upon the change of the ionizing radiation polarization state. CDAD has been used to probe electronic chirality in graphene \cite{Liu11}, helical spin texture in topological insulator \cite{Wang11} and Berry curvature in TMDCs \cite{Cho18,Schuler20}, for example. LDAD is typically assumed to encode the non-relativistic symmetry of the ground state wavefunction \cite{Schonhense90,Cherepkov93,Sterzi18}, which can contain information about the orbital texture \cite{Cao13,Min19}. However, dichroism can also have an extrinsic origin, \textit{i.e.}~it can also emerges from experimental geometry induced symmetry breaking. Disentangling the intrinsic and extrinsic contribution to the dichroic signal is very challenging, but of fundamental importance to extract meaningful physical insight from it.   

Using our multidimensional detection scheme with the $p$-polarized fs-XUV pulses incident in the $k_x$-$k_z$ plane (and along $\Gamma$-M/M'), the normalized intensity differences between the forward ($I_{\alpha}(k_x,k_y,E_B)$) and backward ($I_{\alpha}(-k_x,k_y,E_B)$) hemisphere, \textit{i.e.}~the linear dichroism asymmetry in the photoelectron angular distribution ($A_{LDAD}^{\alpha}(k_x,k_y,E_B)$), can be extracted (see Eq.~\ref{Eq:A_LDAD}), without the need to rearrange the sample geometry or the light-polarization state \cite{Chernov15,Tusche16}.

\begin{equation}
    A_{LDAD}^{\alpha} =
    \frac{I_{\alpha}(k_x,k_y,E_B) - I_{\alpha}(-k_x,k_y,E_B)}{I_{\alpha}(k_x,k_y,E_B) + I_{\alpha}(-k_x,k_y,E_B)}
\label{Eq:A_LDAD}
\end{equation}

Looking at $A_{LDAD}^{0^{\circ}}$ and $A_{LDAD}^{60^{\circ}}$ (Fig.~~\ref{fig2}(d)-(e)), one can notice that some features of the dichroism are invariant upon time-reversal, while others show antisymmetric behavior (sign flip). This can be understood by the fact that the contribution to the dichroism originating from experimental geometry remains unchanged upon 60$\mathrm{^{\circ}}$ rotation of the crystal. Moreover, depending on the energy-momentum region of the electronic structure sampled in each experimental data voxel, the associated ground state wavefunction might be invariant upon time-reversal. For example, the dichroism emerging from the branches pointing along $\Gamma$-M/M' high symmetry direction, which is of multiorbital character ($|p_x\rangle \pm i |p_y\rangle$), switch sign upon crystal rotation. On the other hand, the dichroism emerging from the hexagonal-shaped band surrounding the $\Gamma$-point, which is mostly of single orbital character ($|p_z\rangle$), does not. 

The new observable that we introduce, called time-reversal dichroism in the photoelectron angular distributions (TRDAD), allows to isolate the antisymmetric part of the $A_{LDAD}$ dichroism upon time-reversal, in order to remove any spurious contributions from experimental geometry. TRDAD is defined as, 

\begin{equation}
    TRDAD = \frac{A_{LDAD}^{\alpha} -  A_{LDAD}^{\alpha'}}{2}
    \label{Eq:A_LDAD_T}
\end{equation}

where crystal rotation by an angle $\alpha$-$\alpha'$ ($\mathcal{R}_{\alpha-\alpha'}$) is equivalent to time-reversal, \textit{i.e.}~$\mathcal{R}_{\alpha-\alpha'} \equiv \hat{T}$. The resulting TRDAD is shown in Fig.~\ref{fig2}(f). The branches pointing along $\Gamma$-M have opposite $A_{LDAD}$ with respect to adjacent valleys and dominates the signal, while the signature of the hexagonal-shaped band surrounding the $\Gamma$-point has disappeared. Indeed, a non-vanishing TRDAD signal implies that the state-resolved dichroism changes sign upon time-reversal operation, which we interpret as a switch of the orbital texture.

To investigate the microscopic origin of TRDAD, we perform state-of-the-art quantitative one-step photoemission calculations based on fully relativistic density functional theory (DFT). The one-step model of photoemission is implemented in the fully relativistic Korringa-Kohn-Rostoker (KKR) method. The calculated photoemission signal is layer-resolved and includes all matrix element effects such as experimental geometry, photon energy, polarization state, and final state effects (see SM and refs.~\cite{Ebert11,Braun18}). 

\begin{figure}
\begin{center}
\includegraphics[width=8.5 cm,keepaspectratio=true]{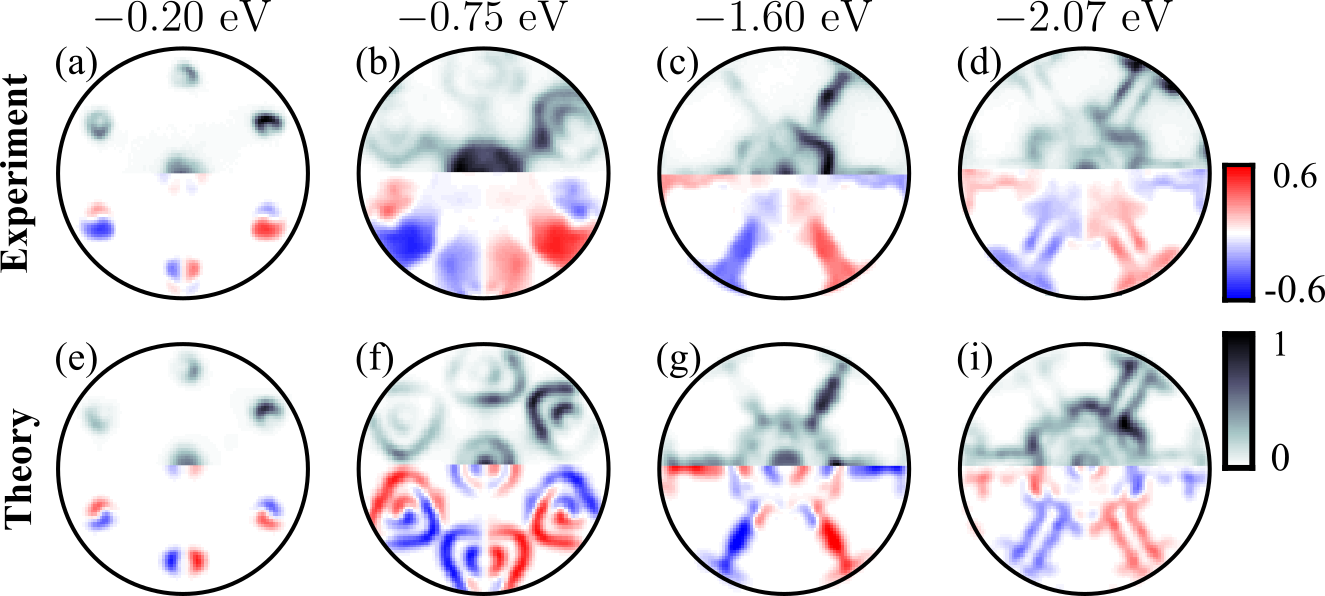}
\caption{\textbf{Comparison between experimentally measured and theoretically (KKR) calculated  TRDAD}. In the upper hemisphere of each panel, the raw photoemission intensity, and in the lower hemisphere, TRDAD. \textbf{(a)-(d)} Experimentally measured photoemission intensity and TRDAD. \textbf{(e)-(i)} Calculated photoemission intensity and TRDAD. \textbf{(a)} and \textbf{(e)} $\mathrm{E-E_{VBM}}$ = -0.20~eV, \textbf{(b)} and \textbf{(f)} $\mathrm{E-E_{VBM}}$ = -0.75~eV, , \textbf{(c)} and \textbf{(g)} $\mathrm{E-E_{VBM}}$ = -1.60~eV and \textbf{(d)} and \textbf{(i)} $\mathrm{E-E_{VBM}}$ = -2.07~eV. The radius of each constant energy contour corresponds to 1.6~{\AA}$^{-1}$.}
\label{fig3}
\end{center}
\end{figure}

\begin{figure}
\begin{center}
\includegraphics[width=8.5 cm,keepaspectratio=true]{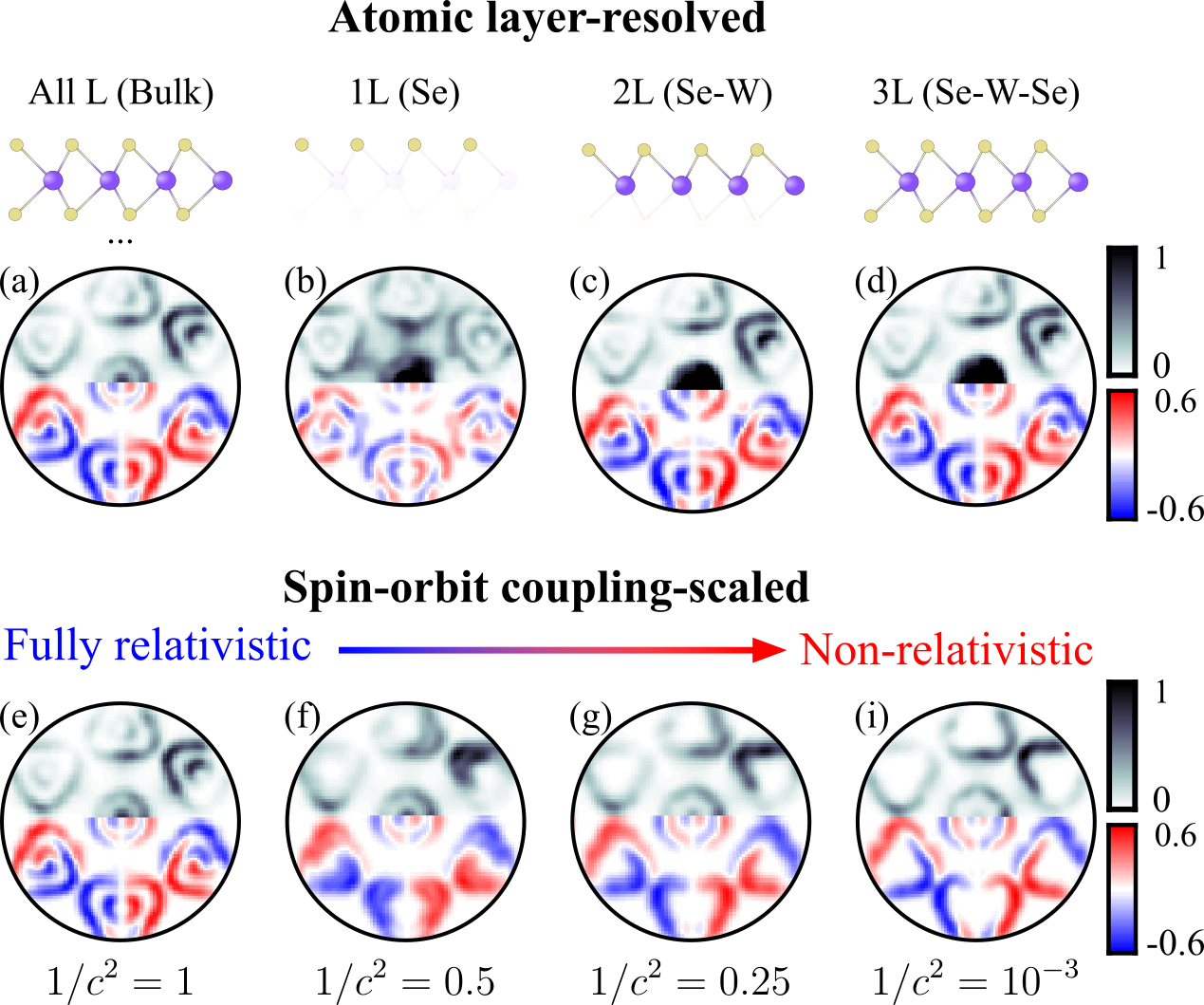}
\caption{\textbf{Theoretical (KKR) investigations of the microscopic origin of the time-reversal dichroism }. In the upper panels, we study the atomic-layer resolved photocurrent and associated TRDAD. The signal is coming from \textbf{(a)} all layers (All L), \textbf{(b)} first Se atomic layer (Se), \textbf{(c)} first Se and W atomic layers (Se-W), and \textbf{(d)} first Se and W, and second Se atomic layers (Se-W-Se). In the lower panels, we investigate the role of spin-orbit coupling on the photocurrent and associated TRDAD, by going from the fully relativistic to the non-relativistic limit, upon modulating the speed of light in the calculations. \textbf{(e)} For 'standard' speed of light (fully relativistic), which we defined as 1/$c^2$=1. \textbf{(f)-(i)} For enhanced speed of light, 1/$c^2$=0.5, 1/$c^2$=0.25, and 1/$c^2$=$\mathrm{10^{-3}}$ (non-relativistic limit), respectively. All the constant energy contours are taken at $\mathrm{E-E_{VBM}}$ = -0.75~eV, and their radii correspond to 1.6~{\AA}$^{-1}$. }
\label{fig4}
\end{center}
\end{figure}

The striking similarity between experimental and theoretical results (Fig.~\ref{fig3}) confirms that the KKR method accurately describes the ground state properties of bulk 2H-WSe$_2$ in an extended binding energy range and captures well subtleties of the photoemission process, including multiorbital interference effects. Now that the ability of the KKR method to quantitatively reproduce the experimentally measured signals is established, we want to strengthen our assert that TRDAD in XUV photoemission probes hidden physical quantities, \textit{i.e.}~quantities that are non-vanishing in each constituent layers but that are vanishing in its inversion-symmetric counterpart (bulk). To do so, we investigated the atomic-layer-resolved photocurrent and associated TRDAD. In Fig.~\ref{fig4}(a), the photoemission intensity and associated TRDAD emerging from all layers (bulk) are presented. The outer (VB1) and inner (VB2) bands around each K/K' valleys show very similar dichroism, \textit{i.e.}~the same positive/negative (red/blue) TRDAD patterns. While the TRDAD signal coming from the topmost selenium (Se) atomic layer (Fig.~\ref{fig4}(b)) is strongly different from the full calculation (Fig.~\ref{fig4}(a)), including the photocurrent from the first tungsten (W) layer (Fig.~\ref{fig4}(c)) is already enough to almost perfectly reproduce all features of the full calculation, which is in good agreement with the predicted W $d$-type orbitals nature of the valence band at K/K' points. These calculations unambiguously confirm that TRDAD probes hidden physical quantities, which are modulated upon time-reversal. 

Next, we want to fully disentangle the signatures of spin and orbital textures in TRDAD. To do so, we have investigated the photoemission intensity and associated TRDAD in both the fully relativistic and non-relativistic limit (vanishing spin-orbit coupling (SOC)). Indeed, SOC is at the origin of the hidden spin-polarization of the two topmost valence bands at K/K', and thus of the emergence of the peculiar spin texture in 2H-TMDCs. In the limit where SOC vanishes (non-relativistic limit), the two oppositely spin-polarized topmost valence bands at K/K' are merging together, leading to the annihilation of the spin-polarization, but to a conservation of the orbital texture. Because the dominant relativistic corrections scale with 1/$c^2$, where $c$ is the speed of light, a straightforward theoretical approach to go from the fully relativistic to the non-relativistic limit, and thus to modify the strength of SOC, is to modulate the speed of light. In Fig.~\ref{fig4}(e), one can see that in the fully relativistic case, the topmost valence band at K/K' are spin-orbit-split, and have similar TRDAD patterns, already suggesting its sensitivity to orbital texture. Strikingly, the non-relativistic TRDAD from the degenerate band at K/K' is identical to the dichroism of the associated topmost valence band in the fully relativistic case. This observation is a smoking-gun evidence that TRDAD is a powerful probe of the hidden orbital texture, which exists even in the absence of spin-orbit coupling and thus of spin texture. 

\begin{figure}
\begin{center}
\includegraphics[width=8.5 cm,keepaspectratio=true]{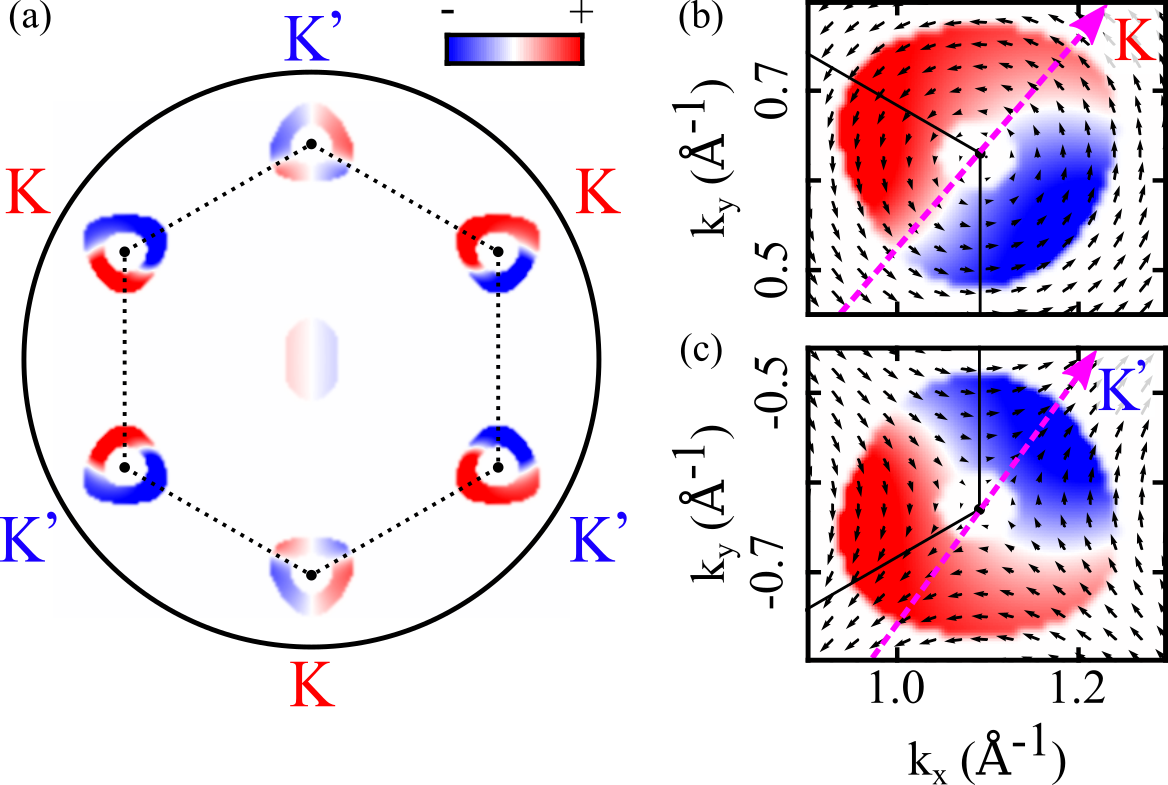}
\caption{\textbf{Linking TRDAD and orbital pseudospin texture using a tight-binding framework}. \textbf{(a)} TRDAD for an energy $\mathrm{E-E_{VBM} = - 0.20~eV}$ computed using the tight-binding formalism described in the SM. \textbf{(b)}, \textbf{(c)} TRDAD and in-plane orbital pseudospin (black arrows) around the K and K' valleys, respectively. The magenta dashed arrow represents the orientation of the orbital vector field, which can be derived from the photoemission matrix elements (more details can be found in the SM). As shown in Eq.~7 of the SM, when the pseudospin texture is parallel/anti-parallel to the orbital vector field, we get a positive/negative signal, while TRDAD vanish if they are orthogonal. This relationship provides a direct link between the experimentally measured TRDAD and the in-plane orbital pseudospin.}
\label{fig5}
\end{center}
\end{figure}

Last, to get a more intuitive and comprehensive picture of the origin of TRDAD, we employ a third-nearest neighbor tight-binding (TB) model \cite{Liu13}, which provides an excellent description of the electronic structure of 2H-WSe$_2$ close to the K/K' points. As shown in Fig.~\ref{fig5}(a), the TB model, for energies close to the valence band maximum, qualitatively reproduces the measured TRDAD. Our TB analysis unambiguously confirms that TRDAD originates from interference between $|d_{z^2}\rangle$ and $|d_{\pm 2}\rangle$ orbitals, which is characterized by the orbital pseudospin texture (see SM). Moreover, as shown in Fig.~\ref{fig5}(b)-(c), TRDAD is directly linked to the projection of the in-plane orbital pseudospin (black arrows) along the direction of the orbital vector field, which can be derived from the photoemission matrix elements (magenta arrows, see SM for details). Indeed, when the orbital pseudospin and the magenta arrow are parallel (anti-parallel) TRDAD is positive (negative), while it vanishes when they are orthogonal. Our tight-binding analysis thus provides a direct relationship between the experimentally measured TRDAD and the orbital pseudospin texture, a quantity of paramount importance in the emergence of Berry curvature and topological properties of matter. 

In conclusion, we have introduced a novel and fully general robust observable in angle-resolved photoemission spectroscopy called time-reversal dichroism in photoelectron angular distributions (TRDAD), which probes the modulation of the photoemission intensity upon crystal rotation mimicking time-reversal. We have demonstrated that the hidden orbital pseudospin texture of prototypical bulk 2H-WSe$_2$ leaves its imprint onto TRDAD through the multiorbital interference process in photoemission. This robust observable is free of contributions from experimental geometry, and is extremely stable against variation of the photon energy, on the contrary to most conventional dichroic ARPES signals, \textit{e.g.} CDAD (see SM). Similar to the role of spin-resolved ARPES to experimentally elucidate complex momentum-space spin texture, we envision that TRDAD could emerge as the new standard observable to probe peculiar momentum-space orbital texture in complex multiorbital materials. Moreover, the extension of the approach to time-resolved TRDAD experiments is conceptually straightforward and will give access to the orbital texture of excited states and changes of topological properties of out-of-equilibrium states of matter, on ultrafast timescales \cite{Sentef15,Claassen16,DeGiovannini16,Oka19,Schueler20}.

\begin{acknowledgments}
\textbf{Acknowledgments} We thank Jürgen Henk and Karsten Horn for enlightening discussions at the early stage of the project. This work was funded by the Max Planck Society, the European Research Council (ERC) under the European Union’s Horizon 2020 research and innovation program (Grant No. ERC-2015-CoG-682843), the German Research Foundation (DFG) within the Emmy Noether program (Grant No. RE 3977/1) and through the SFB/TRR 227 "Ultrafast Spin Dynamics" (projects A09 and B07). J.S. and J.M. would like to thank CEDAMNF project financed by the Ministry of Education, Youth and Sports of Czech Repuplic, Project No. CZ.02.1.01/0.0/0.0/15.003/0000358. J.B. and H.E. acknowledge financial support by the DFG via the projects Eb 158/32 and Eb 158/36. M.S. thanks the Alexander von Humboldt Foundation for its support with a Feodor Lynen scholarship. S.B. acknowledges financial support from the NSERC-Banting Postdoctoral Fellowships Program.
\end{acknowledgments}

\clearpage

\onecolumngrid

\begin{center}
\LARGE{\textbf{Supplementary Material:\\}}

\hspace{2cm}

\large{\textbf{Revealing Hidden Orbital Pseudospin Texture with Time-Reversal Dichroism in Photoelectron Angular Distributions}}
\end{center}

\renewcommand{\thepage}{S\arabic{page}} 
\renewcommand{\thesection}{S\arabic{section}}  
\renewcommand{\thetable}{S\arabic{table}}  
\renewcommand{\thefigure}{S\arabic{figure}} 

\section{Details about the experimental setup}

The experimental apparatus features a table-top femtosecond XUV beamline coupled to a photoemission end-station. Briefly, a home-built optical parametric chirped-pulse amplifier (OPCPA) delivering 15 W (800 nm, 30 fs) at 500 kHz repetition rate \cite{Puppin15} is used to drive high-order harmonic generation (HHG) by tightly focusing the second harmonic of the laser pulses (400 nm) onto a thin and dense Argon gas jet.  The nonperturbative nonlinear interaction between the laser pulses and the Argon atoms leads to the generation of a comb of odd harmonics of the driving laser, extending up to the 11th order. A single harmonic (7th order, 21.7 eV) is isolated by reflection on a focusing multilayer XUV mirror and propagation through a 400 nm thick Sn metallic filter. A photon flux of up to 2x10$^{11}$ photons/s at the sample position is obtained (110 meV FWHM). The bulk 2H-WSe$_2$ samples (HQ Graphene) were cleaved at room temperature and base pressure of 5x10$^{-11}$ mbar, and handled by a 6-axis manipulator (SPECS GmbH). The photoemission data are acquired using a time-of-flight momentum microscope (METIS1000, SPECS GmbH). This detector allows for simultaneous detection of the full surface Brillouin zone, over an extended binding energy range, without the need to rearrange the sample geometry \cite{Medjanik17}. Concerning the data post-processing,  we use a recently developed open-source workflow \cite{Xian19} to efficiently convert the raw single-event-based datasets into binned calibrated data hypervolumes of the desired dimension (here 120x120x120, corresponding to 0.038~{\AA}$^{-1}$ and 67~meV bin sizes), including axes calibration and artifact corrections (including symmetry distortion corrections \cite{Xian19_2}). The resulting 3D photoemission intensity data have the coordinates I($k_x$, $k_y$, $E_B$). 

\section{Details about the KKR calculations}

As mentioned in the manuscript, our photoemission calculations are based on fully relativistic density functional theory (DFT). The one-step model of photoemission is implemented in the fully relativistic Korringa-Kohn-Rostoker (KKR) method of the Munich band structure software package, based on Green’s function and multiple scattering spin-density matrix formalism \cite{Ebert11,Braun18}. The SPR-KKR scheme solves the Dirac equation, hence all the relativistic effects are fully included. The local density approximation (LDA) has been chosen as an exchange-correlation functional. The bulk potential converged in atomic spheres approximation geometry. The bulk 2H-WSe$_{2}$ crystallizes in a $\mathrm{P 6_{3} / m m c D_{6 h}^{4}}$ structure with a lattice constant of 3.280 \r{A}. We employed the same empty sphere placement as described in \cite{Coehoorn87}. To obtain a good fit to the experimental data, it was needed to modify the Wigner-Seitz radius of individual atomic types to the following ratio: W = 1.24 , Se = 1 and the vacuum type = 1.04. We found a good agreement between the ground state potential which we obtained with the ASA and a full potential calculations as implemented in SPR-KKR. We used l$_{max}$=3 to obtain the self-consistent field. After the self-consistency was reached, the one-step model of photoemission was used to calculate the photoemission intensities, using the same geometries as in the experiments. The photoemission signal includes all the matrix element effects such as experimental geometry, photon energy, polarization state, and final state effects.

\section{Band structure: Comparison between experiment and KKR calculation}

\begin{figure}[H]
\begin{center}
\includegraphics[width= 9 cm,keepaspectratio=true]{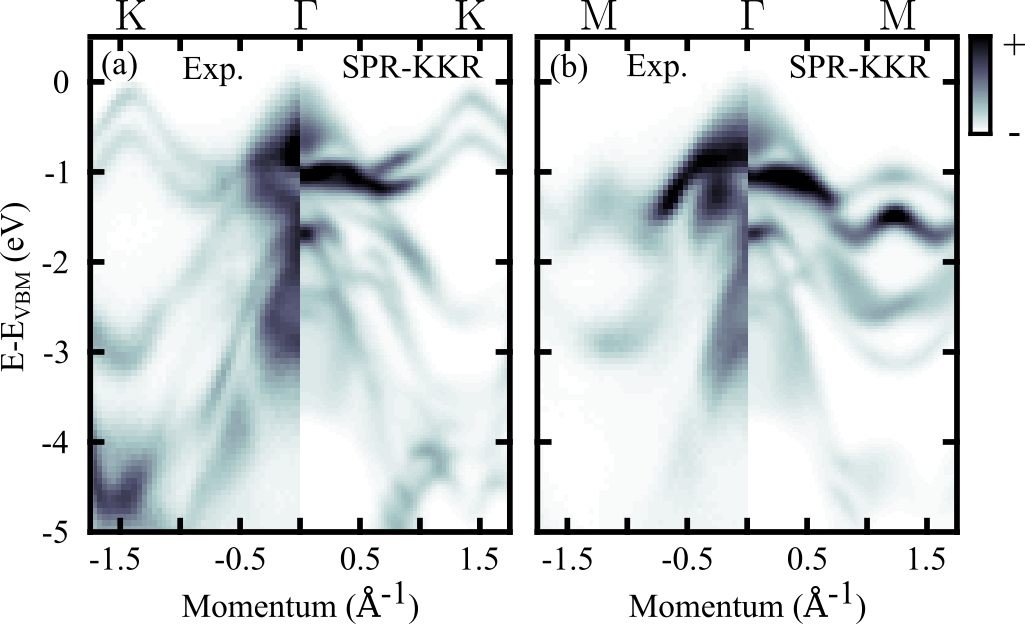}
\caption{\textbf{Energy-momentum cuts along different high symmetry directions: Experiment vs KKR}: \textbf{(a)} The right (left) component of the plot is the photoemission intensity along $\Gamma$-K measured experimentally (calculated using KKR). \textbf{(b)} The right (left) component of the plot is the photoemission intensity along $\Gamma$-M measured experimentally (calculated using KKR).}
\label{Fig_Band_SM}
\end{center}
\end{figure}

As explained in the previous subsection, after the optimization of the Wigner-Seitz radius of individual atomic types, we reached a very good agreement with the experimentally measured electronic band structure of bulk 2H-WSe$_2$. To qualitatively show the agreement between experimentally measured band structure, we have plotted the experimentally measured photoemission intensity along $\Gamma$-K and $\Gamma$-M high symmetry directions, as well as the corresponding calculated photoemission intensity using the one-step model of photoemission (SPR-KKR). The data are shown in Fig.~\ref{Fig_Band_SM}. 

\section{KKR simulations: From the fully relativistic to the non-relativistic limit}

As described in detail in the manuscript, in TMDCs, the combined broken inversion symmetry within each layer and the large spin-orbit coupling lead to peculiar momentum-space spin-orbital textures. To disentangle the effect of spin from orbital degrees of freedom, it would be particularly interesting to have a theoretical way to manipulate the spin-orbit coupling (SOC), and investigate the sensitivity of the time-reversal dichroism in photoelectron angular distributions (TRDAD) to its strength.

For vanishing spin-orbit coupling, the two top-most valence bands, which typically exhibit strong and opposite spin-polarized character around the K/K' points, are expected to merge. This new degeneracy at the Brillouin zone boundary is thus expected to annihilate the spin-polarized character of these bands. However, these two normally spin-split bands have the same orbital texture within each given valley. Thus, the orbital texture is expected to be invariant upon modification of the SOC strength, so is expected to be TRDAD.

Within the KKR framework, there are several ways to manipulate the SOC \cite{Ebert96,Ebert97}. Since SOC is of relativistic origin, and since dominant relativistic corrections scale with 1/$c^2$, where $c$ is the speed of light, one straightforward approach to theoretically mimic the non-relativistic limit is to increase the speed of light. 

\begin{figure}[H]
\begin{center}
\includegraphics[width= 15 cm,keepaspectratio=true]{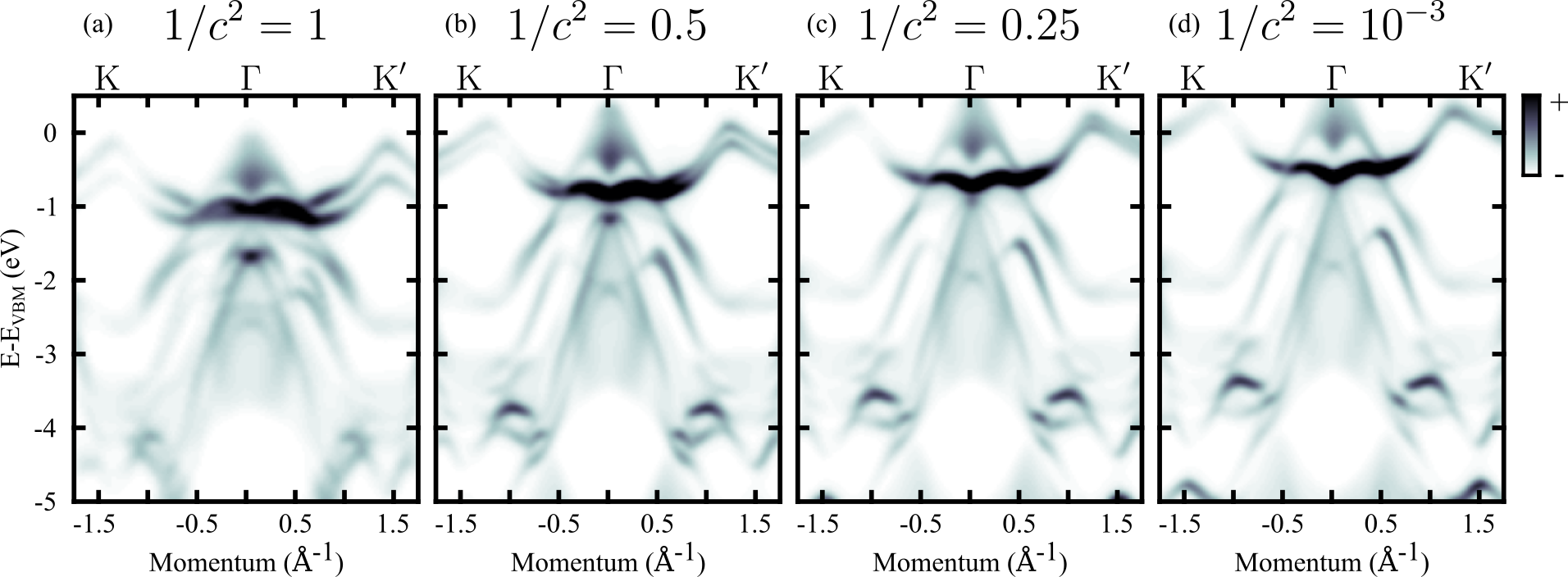}
\caption{\textbf{Energy-momentum cuts along K-$\Gamma$-K': Going from the fully relativistic to the non-relativistic limit}: Calculated photoemission intensity along K-$\mathrm{\Gamma}$-K' high symmetry direction, for different speeds of light. \textbf{(a)} For 'standard' speed of light, which we defined as 1/$c^2$=1. \textbf{(b)-(d)} For enhanced speed of light, 1/$c^2$=0.5, 1/$c^2$=0.25, and 1/$c^2$=$\mathrm{10^{-3}}$, respectively.}
\label{Fig_Band_SOC}
\end{center}
\end{figure}

In Fig.~\ref{Fig_Band_SOC}, we show the calculated band structure along K-$\Gamma$-K' high symmetry direction, for different speeds of light. The same data have been used to extract TRDAD presented in Fig.~4 of the main paper. One can notice that when the speed of light is increased (going from (a) to (d)), the spin-orbit-split band (\textit{e.g.} valence band top at the K/K' points) are merging. 

\begin{figure}[H]
\begin{center}
\includegraphics[width= 11 cm,keepaspectratio=true]{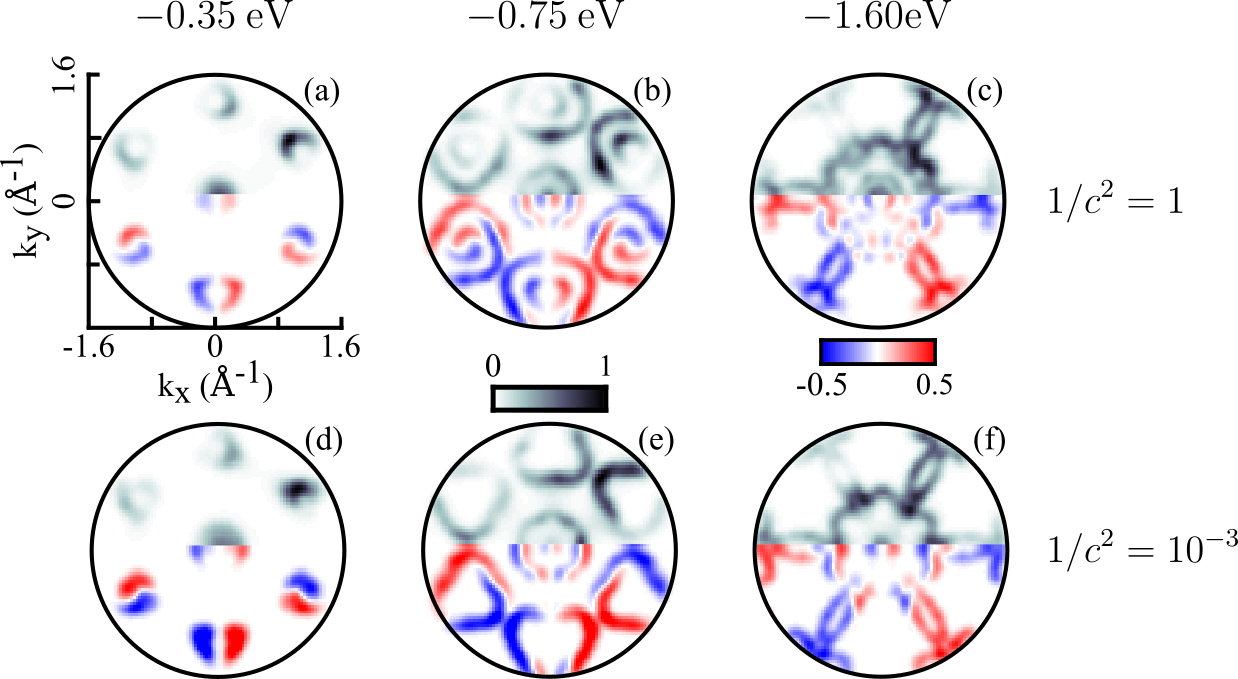}
\caption{\textbf{Fully relativistic and non-relativistic TRDAD}: For all panels, the upper hemisphere is the photoemission intensity for a given crystal orientation, where the scattering plane coincides with the crystal mirror plane (along $\Gamma$-M). The lower hemisphere is the extracted TRDAD. For panels \textbf{(a)-(c)}, we use 'standard' speed of light (1/$c^2$=1), which corresponds to the fully relativistic case. For panels \textbf{(d)-(f)}, we use enhanced speed of light (1/$c^2$=$10^{-3}$), which corresponds to the non-relativistic limit. The energies relative to the valence band maximum are \textbf{(a),(d)} $\mathrm{E-E_{VBM}}$=-0.35 eV, \textbf{(b),(e)} $\mathrm{E-E_{VBM}}$=-0.75 eV and \textbf{(c),(f)}  $\mathrm{E-E_{VBM}}$=-1.60 eV.}
\label{Fig_Light_SM}
\end{center}
\end{figure}

In addition to the Fig.~4 of the manuscript, which shows the effect of the spin-orbit splitting (going progressively from fully relativistic to the non-relativistic limit) on TRDAD for a given binding energy, here we want to present extended data with two different effective speed of light: 1/$c^2$=1 (fully relativistic) and 1/$c^2$=$10^{-3}$ (non-relativistic limit), for different selected binding energies. Fig.~\ref{Fig_Light_SM} (a) and (d) shows that the dichroism slightly below the valence band top ($\mathrm{E_B}$=-0.35 eV) is almost unaffected when going from the fully relativistic to the non-relativistic limit. Indeed, while the absolute amplitude of TRDAD is slightly enhanced in the non-relativistic limit, the alternating positive and negative signal emerging from the croissant shaped photoemission intensity around K/K' valleys is invariant upon modification of the SOC strength. Fig.~\ref{Fig_Light_SM} (b) and (e) shows constant energy contours (CECs) for larger binding energy ($\mathrm{E_B}$=-0.75 eV). In the fully relativistic case (b), one can see that the photoemission intensity features two concentric trigonally warped 'circles' around each K/K' valleys, which can be associated with the two first spin-orbit-split valence bands (VB1 and VB2). The dichroism of both bands is very similar, around each valley, because they have the same hidden orbital texture. When going to the non-relativistic case (e), the inner 'circle' disappear, since the splitting between these bands is of relativistic origin (SOC).  Moreover, one can notice that the dichroism of the outer trigonally warped 'circle' (VB1), does not qualitatively change when going to the non-relativistic case. We can thus conclude that TRDAD is a quantity that is of non-relativistic origin. 

\section{Orbital-resolved TRDAD from KKR calculations}

One other knob that is available within our KKR framework is to turn-off some initial- and final-state channels. We will use this knob to strengthen our conclusions about the microscopic origin of TRDAD. 

\begin{figure}[H]
\begin{center}
\includegraphics[width= 10 cm,keepaspectratio=true]{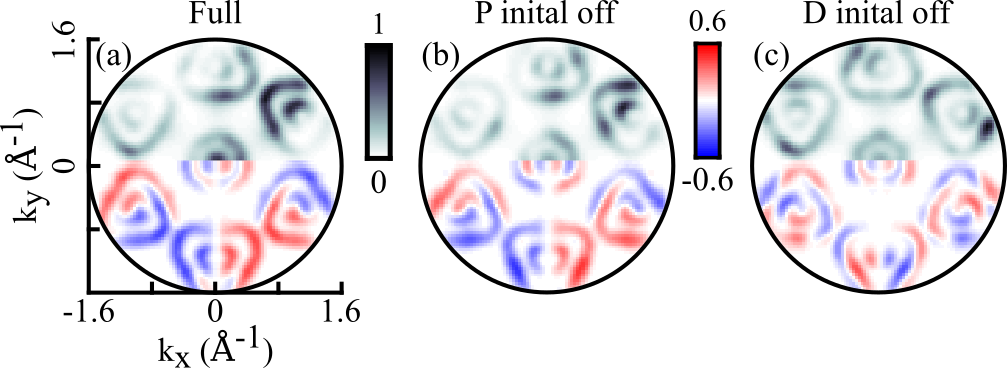}
\caption{\textbf{Orbital-resolved TRDAD from KKR calculations: Role of initial states}: For all panels, the upper hemisphere is the photoemission intensity for a given crystal orientation, where the scattering plane coincides with the crystal mirror plane (along $\Gamma$-M), for a constant energy of $\mathrm{E-E_{VBM}}$ = -0.75 eV. The lower hemisphere is the associated TRDAD. \textbf{(a)} For full calculation. \textbf{(b)} For disabled \textit{p-}type initial states. \textbf{(c)} For disabled \textit{d-}type initial states. }
\label{Fig_Orbitals_Ini_SM}
\end{center}
\end{figure}

We first investigated the role of initial-state channels in the emergence of TRDAD. In Fig.~\ref{Fig_Orbitals_Ini_SM}(b), we show the photoemission intensity and associated TRDAD for the full calculation, \textit{i.e.} where all orbital-types are 'enabled'. In Fig.~\ref{Fig_Orbitals_Ini_SM}(b) and (c), we show the photoemission intensity and associated TRDAD when \textit{p}-type and \textit{d}-type initial orbitals have been disabled in the calculation, respectively. While turning off the contribution of \textit{p}-type orbitals leave TRDAD most unchanged (compared to full calculation), turning off the contribution of \textit{d}-type orbitals leads to a completely different TRDAD signal. These observations confirm our conclusion that TRDAD, for this binding energy, emerges as a consequence of interference between \textit{d-}type orbitals in the photoemission process, as well as the modification of the interferometric pattern upon time-reversal.

\begin{figure}[H]
\begin{center}
\includegraphics[width= 10 cm,keepaspectratio=true]{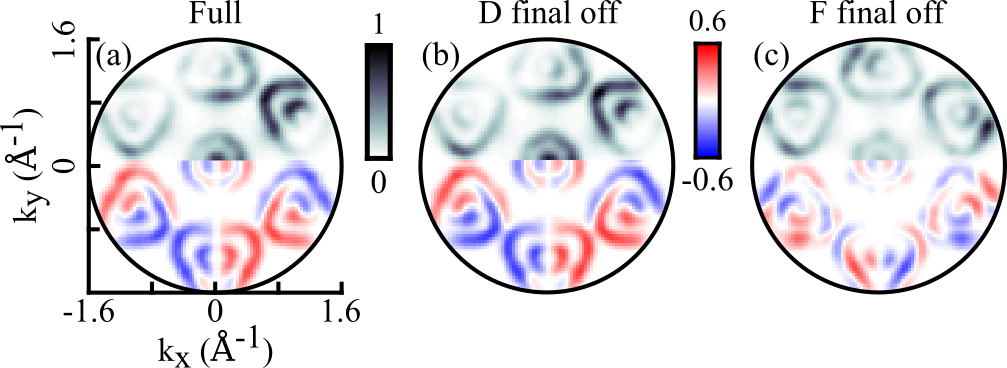}
\caption{\textbf{Orbital-resolved TRDAD from KKR calculations: Role of final states}: For all panels, the upper hemisphere is the photoemission intensity for a given crystal orientation, where the scattering plane coincides with the crystal mirror plane (along $\Gamma$-M), for an constant energy of $\mathrm{E-E_{VBM}}$ = -0.75 eV. The lower hemisphere is the associated TRDAD. \textbf{(a)} For full calculation. \textbf{(b)} For disabled \textit{d-}type final states. \textbf{(c)} For disabled \textit{f-}type final states.}
\label{Fig_Orbitals_Fin_SM}
\end{center}
\end{figure}

Similarly to the above-described procedure, we also investigated the role of final-state channels in the emergence of TRDAD. As one can see in Fig.~\ref{Fig_Orbitals_Fin_SM}, turning off the contribution of \textit{d}-type final states leave TRDAD mostly unchanged (compared to full calculation), while turning off the contribution of \textit{f}-type final states leads to a completely different TRDAD signal. These observations further confirm our conclusion that TRDAD, for this binding energy, emerges as a consequence of interference between \textit{d-}type orbitals in the photoemission process.\\

\section{Photon energy dependence of TRDAD and CDAD from KKR calculations}

The emergence of circular dichroism in photoelectron angular distributions (CDAD), which is a well-established observable in ARPES, can be of multiple origins. For example, photoemission from the helical Dirac fermions at the surface of the topological insulator $\mathrm{Bi_2Te_3}$ leads to strong CDAD, which has been initially interpreted as originating from either the handedness of the experimental setup, the initial-state spin angular momentum, and the initial-state orbital angular momentum. However, a joint experimental and theoretical (KKR calculations, as used in our study) paper demonstrated strong modulation (even sign reversal) of the CDAD signal when scanning the photon energy of the ionizing radiation, and conclude that, in this material, circular dichroism was originating from so-called 'final-state effects' \cite{Scholz13}.

\begin{figure}[t]
\begin{center}
\includegraphics[width= 10 cm,keepaspectratio=true]{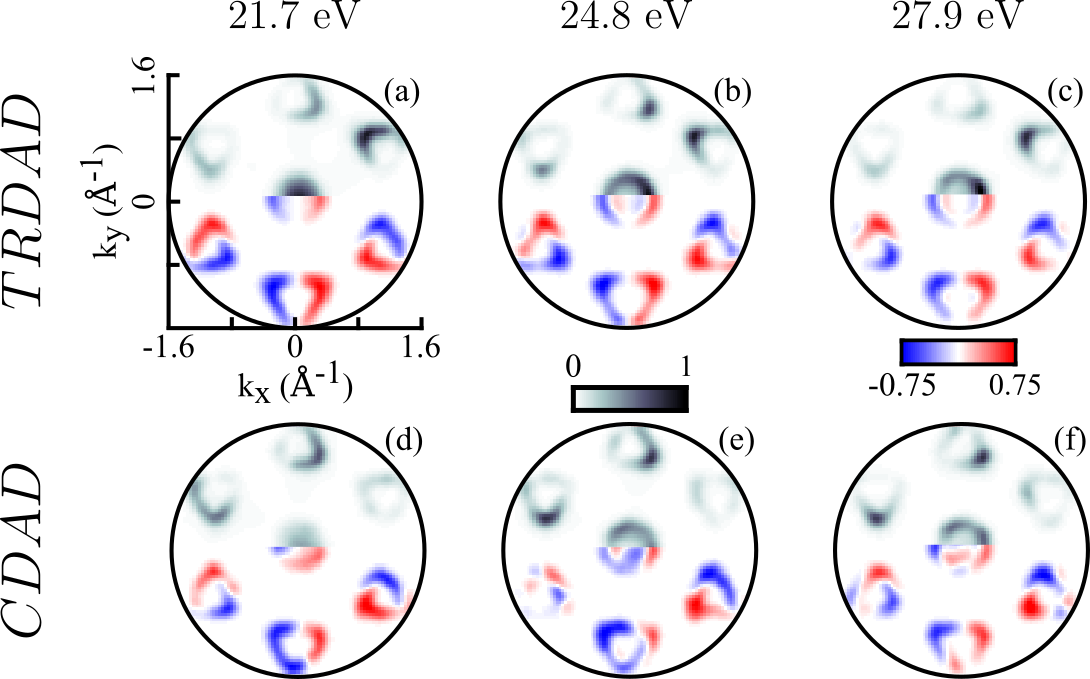}
\caption{\textbf{Comparison of the stability of TRDAD and CDAD against modification of the photon energy, from KKR calculations.} \textbf{(a)-(c)} Photon energy dependence of TRDAD. \textbf{(d)-(f)} Photon energy dependence of CDAD. We used the same photon energy as in the experiments (harmonic 7 of the 3.1 eV driver, \textit{i.e.} 21.7 eV), as well as the photon energies of the two upper neighboring harmonics (\textit{i.e.} 24.8 eV and 27.9 eV).}
\label{Fig_hv}
\end{center}
\end{figure}

That being said, one important question is the following: How stable is the TRDAD signal against variation of the photon energy? In order to investigate this, we have computed (using the KKR framework described in the manuscript and the SM) TRDAD and CDAD for three different photon energies, \textit{i.e.} the same photon energy as in the experiments (harmonic 7 of the 3.1 eV driver, \textit{i.e.} 21.7 eV), as well as the photon energies of the two upper neighboring harmonics (\textit{i.e.} 24.8 eV and 27.9 eV). The results are shown in Fig.~\ref{Fig_hv}.

While TRDAD features are extremely stable against variation of the photon energy, CDAD exhibits strong modulations including even sign reversal upon modification of the photon energy (given our experimental geometry). This unambiguously demonstrated that CDAD is sensitive to details of the experimental parameters/geometry. In particular, the intrinsic dichroism related to the Berry curvature is overshadowed by photon- and geometry-dependent final state effects (using this experimental geometry). On the contrary, our analysis reveals that TRDAD allows extracting the hidden orbital pseudospin texture in an extremely robust fashion. This rules out the predominance of final-state effects in the emergence of TRDAD.

\section{Details about the tight-binding calculations}

To get a more intuitive and comprehensive picture of the origin of TRDAD,  we employ the third-nearest neighbor tight-binding (TB) model from ref.~\cite{liu_three-band_2013}, which provides an excellent description of the electronic structure close to the K, K$^\prime$ points. As the KKR calculations (Fig.~4 in the main text) show, including a monolayer of WSe$_2$ is sufficient to capture the relevant physics. The model includes the $d_{z^2}$, $d_{xy}$ and $d_{x^2-y^2}$ orbitals localized at the W atom. For convenience we rotate to the atomic spherical harmonic basis: $d_0\equiv d_{z^2}$, $d_{\pm 2} \equiv (d_{x^2-y^2} \pm i d_{xy})/\sqrt{2}$, where the index $m$ of orbital $d_m$ corresponds to the angular momentum state $|\ell=2,m\rangle$. We neglect the SOC here, as TRDAD is a predominantly nonrelativistic effect. This is consistent with the TB picture close to K, K$^\prime$, where the SOC gives rise to a spin-dependent shift of the valence band~\cite{fang_textitab-initio_2015}.

The photoemission matrix elements with respect to the valence band can be expressed as
\begin{align}
	\label{eq:matel_orb}
	M(\vec{k},k_\perp) = \sum_{m=0,\pm 2} C_m(\vec{k}) M_m(\vec{k}, k_\perp) \ ,
\end{align}
where $C_m(\vec{k})$ denote the expansion coefficients with respect to the $d_m$ orbital, while $M_m(\vec{k}, k_\perp)$ stands for the corresponding atomic matrix element. For calculating the latter we assume the final states to be plane waves. The dipole operator is expressed in the length gauge, which has been shown to yield qualitatively accurate results~\cite{Schuler20}:
\begin{align}
	\label{eq:matel_atom}
	M_m(\vec{k}, k_\perp) = \int d\vec{r}\, e^{-i\vec{k}\cdot \vec{r}} e^{- i k_\perp z} \vec{u}\cdot \vec{r}\, w_m(\vec{r}) \ .
\end{align}
Here, $\vec{u}$ is the polarization vector, while $w_m(\vec{r})$ is the Wannier function (WF) corresponding to the $d_m$ orbital. We assume the WFs to be well approximated by atomic orbitals: $w_m(\vec{r}) \approx f(r) Y_{\ell=2,m}(\hat{r})$. By using the expansion of plane waves in terms of spherical harmonics, Eq.~\eqref{eq:matel_atom} is efficiently evaluated in terms of the Clebsch-Gordan algebra. To obtain an estimate of the radial dependence $f(r)$, we performed DFT calculations using the {\sc Quantum Espresso} code~\cite{giannozzi_quantum_2009} and computed the projected WFs corresponding to the We $d$ orbitals using the {\sc Wannier90} code~\cite{pizzi_wannier90_2020}. 
We have extracted the radial function $f(r)$ from the thus obtained $d_{z^2}$ orbital. Test calculations with Slater-type or Gaussian-type wave-functions show that inserting a realistic shape of $f(r)$ into eq.~\eqref{eq:matel_atom} is important for capturing the photoemission features. 

\section{Interference and orbital pseudospin analysis from tight-binding calculation}

The TB model allows to qualitatively reproduce the measured TRDAD (see Fig.~5 in the main text). The flexibility and simplicity of the model enable us to unveil the origin of the distinct features of the $\vec{k}$-dependence of the spectra and the resulting TRDAD. 

\begin{figure}[ht]
	\centering
	\includegraphics[width=\textwidth]{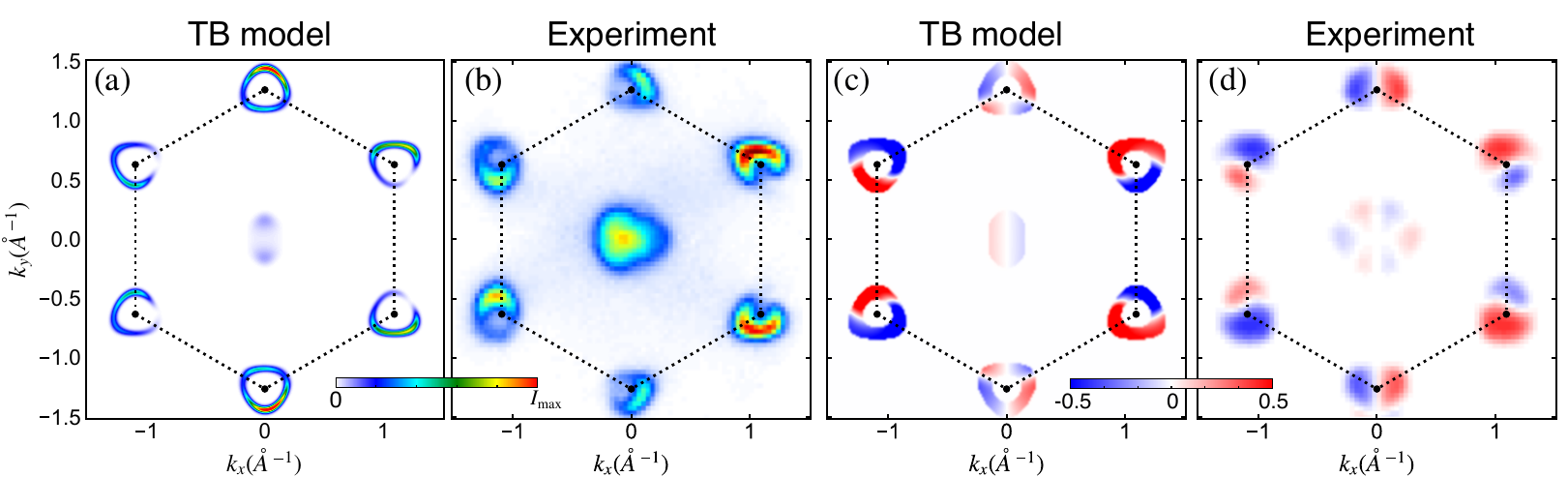}
	\caption{\textbf{Momentum-resolved photoemission intensity and TRDAD from TB model}: \textbf{(a)} Calculated photoemission intensity using the experimental geometry, compared to the \textbf{(b)} experimental spectrum. \textbf{(c)} Calculated TRDAD, while \textbf{(d)} shows the corresponding signal obtained from the experiment. For each panel, $\mathrm{E-E_{VBM}}$ = -0.2~eV.}
  \label{fig:inten_trd_exp_theo}
\end{figure}

Including the $d_{z^2}$ and $d_{\pm 2}$ orbitals in the TB model and calculating the photoemission intensity based on Eq.~\eqref{eq:matel_orb}--\eqref{eq:matel_atom} yields the characteristic spectrum depicted in Fig.~\ref{fig:inten_trd_exp_theo}(a). As in the experiment (Fig.~\ref{fig:inten_trd_exp_theo}(b)), there is a pronounced angular dependence around each K, K$^\prime$ point ("croissant" shape). The position of the "dark" corridor of lower intensity -- in particular, the symmetry -- matches the experiment. Furthermore, the corresponding TRDAD signal with the TB-ARPES model (Fig.~\ref{fig:inten_trd_exp_theo}(c)) is in excellent qualitative agreement with the experiment (Fig.~\ref{fig:inten_trd_exp_theo}(d)).

\begin{figure}[ht]
	\centering
	\includegraphics[width=\textwidth]{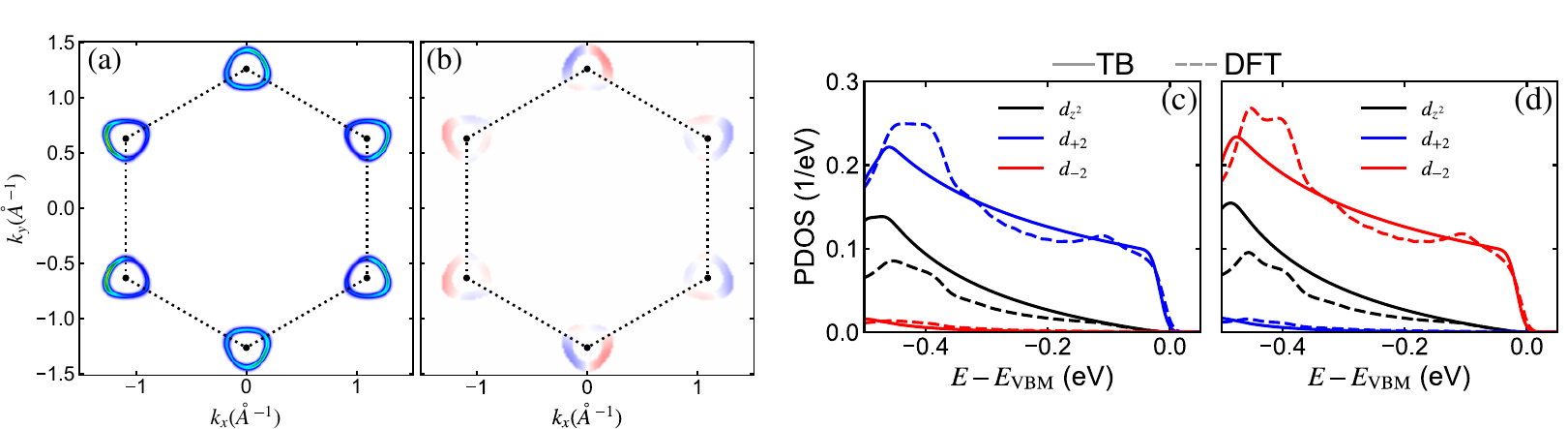}
	\caption{\textbf{Projected density of states around the K (K$^\prime$) valleys and TB momentum-resolved photoemission intensity and TRDAD without $d_{z^2}$ orbitals}: \textbf{(a)} Photoemission spectrum as in Fig.~\ref{fig:inten_trd_exp_theo}(a) excluding the $d_{z^2}$ orbital ($\mathrm{E-E_{VBM}}$ = -0.2~eV). \textbf{(b)}: Corresponding TRDAD signal ($\mathrm{E-E_{VBM}}$ = -0.2~eV). The color scale of \textbf{(a)} (\textbf{b}) is consistent with Fig.~\ref{fig:inten_trd_exp_theo}(a) (Fig.~\ref{fig:inten_trd_exp_theo}(c)-(d)). \textbf{(c)}, \textbf{(d)}: Comparison of the orbital-resolved density of states of the TB model and the projected density of states from the DFT calculation, for K and K' valley, respectively.  \label{fig:inten_trd_nodz2_orbweight}}
\end{figure}

To understand the origin of this intensity variation, we consider the orbital contribution to the photoemission signal. The orbital character close to K (K$^\prime$) is dominated by $d_{+2}$ ($d_{-2}$) (directly at K (K$^\prime$) $d_{+2}$ ($d_{-2}$) is an eigenstate), with a contribution of $d_{z^2}$ increasing further away from K, K$^\prime$, as inferred by inspecting the projected density of states (Fig.~\ref{fig:inten_trd_nodz2_orbweight}(a), (b)). The TB model captures the relative weights of the $d$ orbitals accurately close to K, K$^\prime$.

As a test, we exclude the contribution of the $d_{z^2}$ orbital in Eq.~\eqref{eq:matel_orb}. The resulting photoemission intensity is shown in Fig.~\ref{fig:inten_trd_nodz2_orbweight}(c). Without the $d_{z^2}$ orbital, the intensity variations of the signal around K/K' valleys are much less pronounced. Furthermore, the symmetry of the does not match the experiment. This indicates that the $d_{z^2}$ orbital plays an important role in the observed "croissant" shape. 
Note that even the weight of the $d_{z^2}$ orbital is small in the vicinity of the K, K$^\prime$ points~\cite{fang_textitab-initio_2015}, the large out-of-plane component of the light polarization coalesces in a significant contribution.

This picture is further confirmed by comparing the TRDAD signal including (Fig.~\ref{fig:inten_trd_exp_theo}(c)) and excluding (Fig.~\ref{fig:inten_trd_nodz2_orbweight}(d)) the $d_{z^2}$ orbital, respectively. Without the $d_{z^2}$ contribution, the magnitude of TRDAD is strongly reduced; moreover, the variation from positive to negative values is not consistent with the experiment. In contrast, the three-orbital TB model (Fig.~\ref{fig:inten_trd_exp_theo}(c)) matches the measured TRDAD (Fig.~\ref{fig:inten_trd_exp_theo}(d)) qualitatively very well. This analysis clearly evidences the fundamental role of the $d_{z^2}$ orbitals in the emergence of time-reversal signal.

\begin{figure}[ht]
	\centering
	\includegraphics[width=\textwidth]{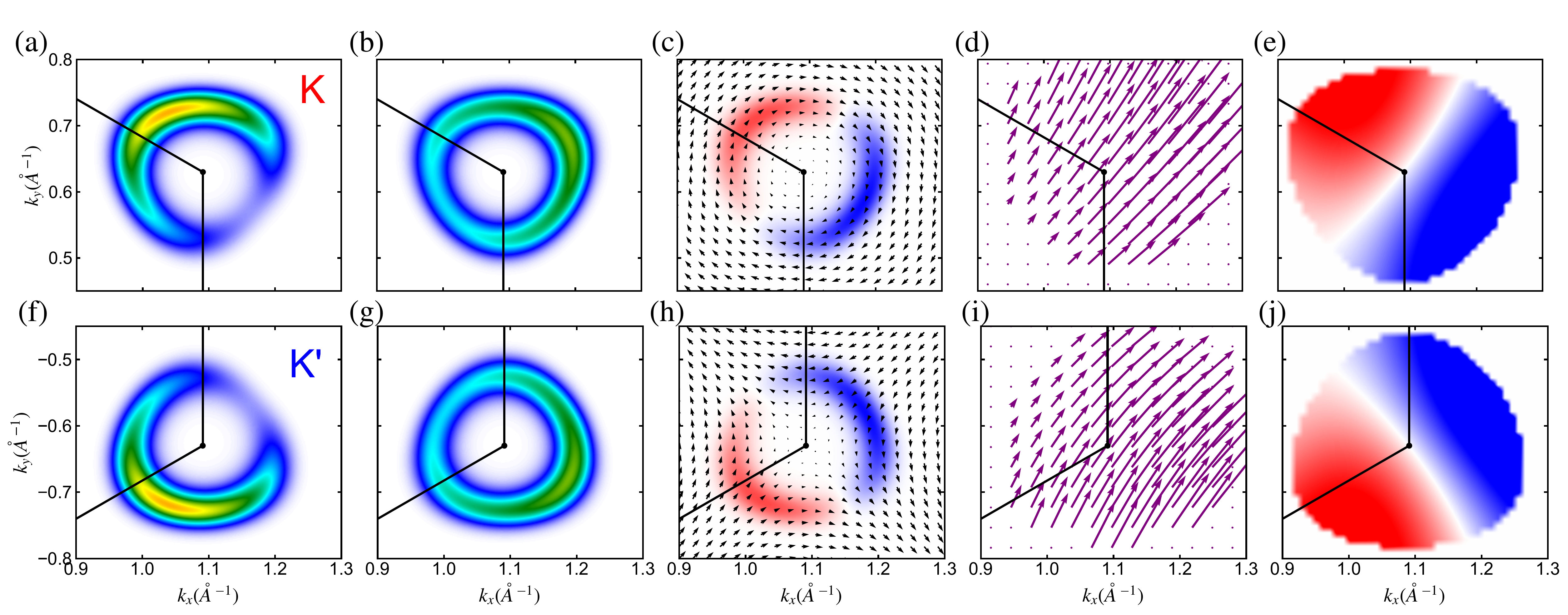}
	\caption{\textbf{Pseudospin analysis close to the valence band maximum}: The upper (lower) panels are showing analysis for a given K (K') valley. \textbf{(a),(f)} Photoemission intensity calculated from the TB model (as in Fig.~\ref{fig:inten_trd_exp_theo}(a)) in the vicinity of K and K', respectively. \textbf{(b),(g)} Incoherent photoemission intensity $I_\mathrm{incoh}(\vec{k},E) $ (excluding $F_\mathrm{int}$ from Eq.~\eqref{eq:inten_K}). 
	\textbf{(c),(h)} Interference contribution to the photoemission intensity $I_\mathrm{int}(\vec{k},E)$ (excluding $F_{0,+2}$ from Eq.~\eqref{eq:inten_K}). Blue (red) color corresponds to suppression (enhancement) of the intensity with respect to \textbf{(b),(g)}.
	The overlaid vector field represents the in-plane orbital pseudospin $\gvec{\sigma}(\vec{k})$. \textbf{(d),(i)} Vector field $\vec{R}(\vec{k})$ according to Eq.~\eqref{eq:rvec}. \textbf{(e),(j)} Scalar product $\vec{R}(\mathrm{K},p_\perp)\cdot \gvec{\sigma}(\vec{k})$. For each panel, the binding energy is $E-E_\mathrm{VBM}=-0.20$~eV.
	\label{fig:pseudo_rdots_dz2}}
\end{figure}

Now we show that the characteristic TRDAD is directly connected to the interference of $d_{z^2}$ and $d_{\pm 2}$ orbitals. Let us consider a K valley, where we can employ a two-orbital model close to the valence band maximum (as the weight of $d_{-2}$ is negligible, see Fig.~\ref{fig:inten_trd_nodz2_orbweight}(c)). The photoemission intensity is then given by
\begin{align}
	\label{eq:inten_K}
	I(\vec{k},E) = \left|M(\vec{k},k_\perp) \right|^2 \delta(E + \hbar \omega - E_k) = \left|C_0(\vec{k}) M_0(\vec{k},k_\perp) + C_{+2}(\vec{k}) M_{+2}(\vec{k},k_\perp) \right|^2 \delta(E + \hbar \omega - E_k)   \ ,
\end{align}
where $E_k = (\vec{k}^2 + k^2_\perp)^2/2$, while $\hbar \omega$ denotes the photon energy. Rearranging the matrix element into
\begin{align}
	\left|M(\vec{k},k_\perp) \right|^2 = F_0(\vec{k},k_\perp) + F_{+2}(\vec{k},k_\perp) + F_{\mathrm{int}}(\vec{k},k_\perp)
\end{align}
with $F_{0,+2}(\vec{k},k_\perp) = |C_{0,+2}(\vec{k})|^2 |M_{0,+2}(\vec{k},k_\perp)|^2$ and 
\begin{align}
	\label{eq:fint}
	F_{\mathrm{int}}(\vec{k},k_\perp) = 2 \mathrm{Re}\left[ C^*_0(\vec{k}) C_{+2}(\vec{k}) M^*_0(\vec{k},k_\perp) M_{+2}(\vec{k},k_\perp)  \right]
\end{align} 
allows for distinguishing the individual orbital contribution and their interference (Eq.~\eqref{eq:fint}). Comparing the full 
photoemission intensity for $E-E_\mathrm{VBM}=-0.20$~eV (Fig.~\ref{fig:pseudo_rdots_dz2}(a),(f)) to the incoherent sum of the orbital contribution $I_\mathrm{incoh}(\vec{k},E) = (F_0(\vec{k},k_\perp) + F_{+2}(\vec{k},k_\perp)) \delta(E + \hbar \omega - E_k)$, presented in Fig.~\ref{fig:pseudo_rdots_dz2}(b),(g), reveals that the characteristic "dark" corridor is diminished if the interference between the orbitals is neglected. The interference contribution $I_\mathrm{int}(\vec{k},E) = F_\mathrm{int}(\vec{k},k_\perp)  \delta(E + \hbar \omega - E_k)$ (Fig.~\ref{fig:pseudo_rdots_dz2}(c),(h)) is pronounced, especially in the direction of the dark corridor. 
To understand the symmetry of the interference effects, we computed the orbital pseudospin
\begin{align}
	\gvec{\sigma}(\vec{k}) = [C^*_0(\vec{k}), C^*_{+2}(\vec{k})] \hat{\gvec{\sigma}} \begin{bmatrix} C_0(\vec{k}) \\ C_{+2}(\vec{k})   \end{bmatrix} \ ,
\end{align}
where $\hat{\gvec{\sigma}} = [\hat{\sigma}_x, \hat{\sigma}_y, \hat{\sigma}_z]$ is the vector of Pauli matrices. The in-plane pseudospin texture is shown in Fig.~\ref{fig:pseudo_rdots_dz2}(c), underlining a direct correlation of the pseudospin texture and the interference contribution. Indeed, Eq.~\eqref{eq:fint} can be expressed as
\begin{align}
	\label{eq:fint_pseudo}
	F_{\mathrm{int}}(\vec{k},k_\perp) = \vec{R}(\vec{k},k_\perp) \cdot \gvec{\sigma}(\vec{k}) \ ,
\end{align}
with
\begin{align}
	\label{eq:rvec}
	R_x(\vec{k},k_\perp) = \mathrm{Re}[M^*_0(\vec{k},k_\perp) M_{+2}(\vec{k},k_\perp)] \ , \ 
	R_y(\vec{k},k_\perp) = -\mathrm{Im}[M^*_0(\vec{k},k_\perp) M_{+2}(\vec{k},k_\perp)]
\end{align}

Inspecting the vector field $\vec{R}(\vec{k},k_\perp)$ close to the K/K' points (Fig.~\ref{fig:pseudo_rdots_dz2}(d),(i)) show an almost uniform behavior. Hence, $F_{\mathrm{int}}(\vec{k},k_\perp)$ is governed by projection of the pseudospin texture along a fixed direction. This is confirmed by approximating $F_{\mathrm{int}}(\vec{k},k_\perp) \approx \vec{R}(\mathrm{K},k_\perp) \cdot \gvec{\sigma}(\vec{k})$, presented in Fig.~\ref{fig:pseudo_rdots_dz2}(e),(j). 

As the analysis reveals, the dark corridor in the photoemission intensity is governed by the in-plane pseudospin projection. This is a genuine interference effect. Since the time-reversal signal is dominated by the observed rotation of the dark corridor upon crystal rotation, the sign structure and the symmetries of the interference term is inherited by TRDAD, thus providing direct insight into the pseudospin texture. 

To illustrate this point, the direction of $\vec{R}(\mathrm{K},k_\perp)$ and $\vec{R}(\mathrm{K}^\prime,k_\perp)$ at the corresponding K, K$^\prime$ points are shown in Fig.~5 in the main text along with the pseudospin texture. The symmetries of the pseudospin projection (Fig.~\ref{fig:pseudo_rdots_dz2}(e),(j)) are in line with the time-reversal signal.


\providecommand{\noopsort}[1]{}\providecommand{\singleletter}[1]{#1}%

\end{document}